\documentclass[aps,prb,twocolumn,showpacs,superscriptaddress,10pt,floatfix]{revtex4-2}
\usepackage{graphicx}
\usepackage{amsmath,amssymb}
\usepackage[utf8]{inputenc}
\usepackage[colorlinks = true,
linkcolor = blue,
urlcolor  = blue,
citecolor = blue,
anchorcolor = blue]{hyperref}
\usepackage{multirow}
\usepackage{braket}
\usepackage{appendix}
\usepackage{color}
\usepackage{times}

\begin{document}
\title{Controlled probing of localization effects in non-Hermitian Aubry-Andr\'e model via topolectrical circuits}
\author{Dipendu Halder}
\email[corresponding author: ]{h.dipendu@iitg.ac.in}

\author{Saurabh Basu}
\email[]{saurabh@iitg.ac.in}
\affiliation{Department of Physics, Indian Institute of Technology Guwahati-Guwahati, 781039 Assam, India}

\begin{abstract}
\noindent Anderson localization and the non-Hermitian skin effect are two distinct confinement phenomena of the eigenfunctions that are, respectively, driven by disorder and non-reciprocity.
Understanding their interplay within a unified framework offers valuable insights into the localization properties of low-dimensional systems.
To this end, we investigate a non-Hermitian version of the celebrated Aubry-Andr\'e model, which serves as an ideal platform due to its unique self-dual properties and ability to demonstrate delocalization-localization transition in one dimension.
Interestingly, in our setting, the competition between Anderson localization and skin effect can be precisely controlled via the complex phase of the quasiperiodic disorder.
Additionally, by analyzing the time evolution, we demonstrate that quantum jumps between the skin states and the Anderson-localized states occur in the theoretical model.
Further, to gain support for our theoretical predictions in an experimental platform, we propose a topolectrical circuit featuring an interface that separates two distinct electrical circuit networks.
The voltage profile of the circuit exhibits confinement at the interface, analogous to the skin effect, while the phenomenon of Anderson localization in the circuit can be perceived via a predicted localization behavior near the excitation node, rather than exhibiting sudden non-Hermitian jumps, as observed in the tight-binding framework.
This interplay leads to a spatially tunable localization of the output voltage of the circuit.
Our findings provide deeper insights into the controlled confinement of the eigenstates of the non-Hermitian Aubry-Andr\'e model by designing analogous features in topolectrical circuits, opening avenues in the fabrication of advanced electronic systems, such as highly sensitive sensors and efficient devices for information transfer and communication.
\end{abstract}

\maketitle
\section{\label{s1}Introduction}
Disorders, impurities, and defects are inherent properties of material preparation.
A particularly intriguing phenomenon linked to disorder in condensed matter physics is the Anderson localization \cite{PhysRev.109.1492} (AL) describes how an infinitesimal random disorder induces a transition from an extended to a localized phase in a system in any dimension less than three.
Interestingly, AL is not limited to systems with random disorder; quasiperiodic (QP) disorders with incommensurate periods can also result in AL.
Among the various QP models, the Aubry-Andr\'e (AA) model \cite{Aubre} has garnered significant attention for its theoretical elegance \cite{PhysRevLett.110.180403, PhysRevLett.114.146601, PhysRevLett.118.016804} and experimental realizations in platforms like photonic crystals \cite{PhysRevLett.103.013901, PhysRevLett.109.106402, DEY}, ultra-cold atoms \cite{Roati2008, PhysRevLett.126.040603}, and superconducting circuits \cite{Li2023}.
A hallmark of the AA model is the absence of mobility edges, an energy-dependent localization transition in the system.
This robustness makes the AA model an excellent platform for exploring localization phenomena.

In recent years, non-Hermitian (NH) phenomena have experienced remarkable growth in a parallel framework, finding applications across a wide range of condensed matter systems \cite{PhysRevLett.116.133903, PhysRevLett.120.146402, Ashida2020, RevModPhys.93.015005}.
This exciting area has unveiled a wealth of novel physical phenomena, such as the non-Hermitian skin effect (NHSE) \cite{PhysRevLett.121.086803, PhysRevB.99.201103, PhysRevLett.124.086801, PhysRevLett.124.056802, Zhang, Okuma}, where the bulk eigenstates accumulate near the boundaries, and the emergence of exceptional points \cite{Heiss_2012, PhysRevLett.118.040401}, where the Hamiltonian becomes defective with the eigenvalues and the eigenvectors being coalesced.
Additionally, the non-Bloch band theory \cite{PhysRevLett.121.086803, PhysRevLett.123.066404} has redefined the conventional Bloch theorem, offering new insights into the wave behavior of NH systems.
Experimental advancements have validated these phenomena in diverse physical scenarios, including ultra-cold atoms \cite{Eichelkraut2013, El-Ganainy2018}, mechanical \cite{Wang}, photonic \cite{Weidemann2020}, and acoustic systems \cite{Fleury2015, 10.1063/5.0186638, 10.1063/5.0237506}.
These developments have established NH systems as apt avenues for exploring the interplay between topology and non-hermiticity.

Furthermore, the interplay between disorder and non-hermiticity has also gained significant attention, particularly with the proposal of the Hatano-Nelson model in 1996 \cite{PhysRevLett.77.570}.
Through the tight-binding (TB) framework, this one-dimensional (1D) model, characterized by asymmetric hopping and random disorder, reveals the delocalization-localization (DL) transition.
Subsequently, disordered NH systems, particularly NH QP models \cite{PhysRevLett.122.237601, PhysRevA.103.L011302, PhysRevB.103.054203, PhysRevB.104.024201, PhysRevB.106.214207, PhysRevB.108.014204, PhysRevB.110.054202, PhysRevB.110.134203}, have attracted significant research interest.
Additionally, the study of localization and topological phase transitions in generalized NH AA models with incommensurately modulated asymmetric hopping amplitudes offers valuable insights \cite{PhysRevB.106.214207, PhysRevB.108.014204}.
Among various experimental platforms, topolectrical circuits (TECs) have evolved as a powerful tool in experiments, drawing attention for their ability to map TB Hamiltonians onto circuit Laplacians \cite{PhysRevLett.114.173902, Lee2018, PhysRevB.99.161114, PhysRevResearch.3.023056, 10.1063/5.0157751, YANG20241}.
By adjusting electrical components and connection configurations, TECs offer remarkable flexibility to engineers and explore a wide range of topological characteristics.
In TECs, topological edge states are revealed through the impedance or voltage profiles, which can be measured by exciting specific nodes within the circuit network.
These features make TECs an excellent medium for studying both theoretical and experimental aspects of topology in condensed matter systems.

While a substantial body of literature exists on the realization of the NHSE using TECs \cite{Helbig2020, PhysRevResearch.5.043034, 10.1063/5.0230976, PhysRevB.109.115407, Guo2024, rafiul} and on AL in QP systems \cite{PhysRevB.100.054301, PhysRevResearch.2.033052, PhysRevB.105.014207, Ganguly2023, PhysRevB.108.144203}, the intricate interplay between these two phenomena has been addressed in only a few recent studies \cite{SUN2023129043, PhysRevB.100.054301, PhysRevB.111.L140201, PhysRevB.110.174203}.
For instance, Ref.~\cite{PhysRevB.110.174203} explores a Hatano–Nelson model with coexisting QP and periodic potentials and proposes a TEC realization.
However, it does not demonstrate the competition between AL and NHSE within the circuit itself.
Other theoretical works, such as Refs.\cite{SUN2023129043, PhysRevB.111.L140201}, analyze this competition, but do not address any experimental implementation.
In contrast, Refs.~\cite{PhysRevResearch.5.043034, Guo2024} present experimental demonstrations of phenomena like higher-order skin effects and scale-tailored localization in TECs, albeit in the absence of any QP disorder.
Similarly, while Ref.~\cite {PhysRevB.100.054301} investigates the AL–NHSE transition in a non-reciprocal AA model and proposes a TEC framework; it does not provide a means to control the interplay between these two localization mechanisms, either in the TB model or its circuit realization.
In this work, we bridge these gaps by offering a comprehensive analysis of the competition between AL and NHSE by demonstrating its realization in a classical, experimentally accessible platform.

In this work, we examine the competition between the AL and the NHSE, along with their time evolution, proposing a design of a TEC that serves as a direct classical analog of these quantum localization phenomena.
Starting with a 1D NH AA model featuring an interface separating two non-equivalent AA chains, we notice intriguing phenomena arising from the interplay between AL and NHSE.
Utilizing standard circuit elements, we construct a TEC capable of replicating both the Hermitian and NH versions of the AA model.
Hence, by exciting a random node with an external source, we successfully observe the classical analogs of both AL and NHSE.
A striking phenomenon emerges when these two compete in a TEC, providing the ability to precisely manipulate the localization of the voltage profile (VP) to specific nodes or ranges within the circuit network, as well as the ability to control the amplitude of the VP.

The paper is organized as follows:  
Section~\ref{s2} introduces the theoretical TB framework, outlining the fundamental physics.  
In section~\ref{s3}, we investigate the time evolution of the AA model when excited at an arbitrary site.  
Section~\ref{s4} explores the electrical analogs of AL and NHSE in the TEC through voltage measurements performed using LTspice software by {\it Analog Devices} \cite{LTspice}.
This section also examines the interplay between AL and NHSE within the TEC.  
Finally, section~\ref{s5} summarizes our findings and discusses potential experimental realizations or device implementations inspired by our theoretical framework.

\section{\label{s2}Theoretical model}
We adopt the non-reciprocal version of the NH AA model introduced by S. Longhi \cite{PhysRevLett.122.237601}, incorporating an interface at a particular lattice site of the chain.
The Hamiltonian (in 1D) is given by,
\begin{widetext}
\begin{equation}
H=\sum_{k=1}^{L_0}\left[(t+\gamma)\hat{c}^{\dagger}_{k+1}\hat{c}_{k}+(t-\gamma)\hat{c}^{\dagger}_{k}\hat{c}_{k+1}\right]+ 
\sum_{k=L_0+1}^{2L}\left[(t-\gamma)\hat{c}^{\dagger}_{k+1}\hat{c}_{k}+(t+\gamma)\hat{c}^{\dagger}_{k}\hat{c}_{k+1}\right] + 
\sum_{k=1}^{2L+1}\lambda_k\hat{c}^{\dagger}_{k}\hat{c}_{k};\quad \lambda_k=2\lambda\cos (2\pi\beta k+i\alpha),
\label{eq:Ham}
\end{equation}
\end{widetext}
where $t$, $\gamma$, and $\lambda$ denote the strengths of the nearest-neighbor hopping, non-reciprocity, and QP disorder, respectively.
All of these parameters are assumed to be real and positive.
Note that all parameters in this TB model are in the unit of $t$.
\begin{figure}
    \includegraphics[width=\columnwidth]{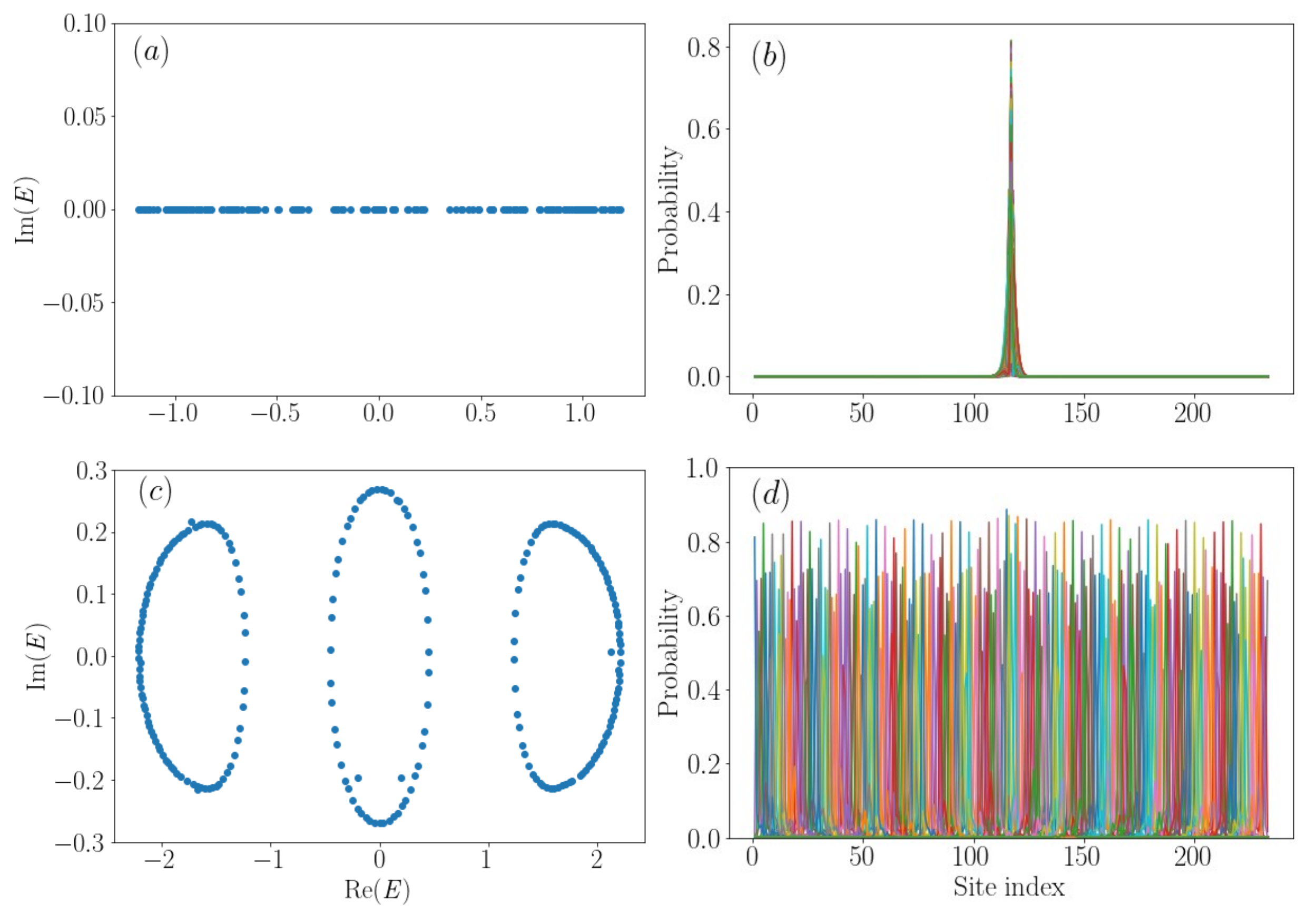}
    \caption{Numerical analyses for a total number of lattice sites, $(2L+1) = 233$, being a Fibonacci number, with $t=0.65$ and $\gamma=0.35$, are presented. (a) Real vs Imaginary parts of the energy spectra for $\lambda=\alpha=0$; (b) Probability distribution of the corresponding eigenstates, which localize at site index $(L+1) = 117.$ (c) Same as (a) but with $\lambda=0.9,\,\alpha=0.2.$ (d) The skin states in (b) have now become AL states as $\alpha>\alpha_c\simeq 0.105.$}
    \label{fig:1}
\end{figure}
Our system comprises of $(2L+1)$ lattice sites, and $(L_0+1)^{\text{th}}$ site marks the position of the interface, starting from the first node.
For concreteness, we set $L_0 = L$ in Eq.~\eqref{eq:Ham} for further discussions and the corresponding circuit design.
The parameter $\alpha$ introduces a complex component to the QP potential and plays a pivotal role in the delocalization-localization transition of our model.
Its prominence will become evident in the subsequent discussions.
$\beta$ is an irrational number given by $\beta=(\sqrt{5}-1)/2$ with $\beta^{-1}$ being the golden ratio.
It is obtained via $\beta= \lim_{n\to\infty} \left(\frac{F_{n-1}}{F_n}\right),$ where the Fibonacci numbers $F_n$s are defined recursively by $F_{n+1}=F_n+F_{n-1}$ and $F_0 = 0, F_1 = 1.$
The operators $\hat{c}_{k}$ and $\hat{c}_{k}^{\dagger}$ denote the annihilation and creation operators for spinless fermions at the site $k$.

The behavior of the system depends on two distinct scenarios: (a) $\lambda = 0$: in the absence of the QP potential, the system reduces to the clean Hatano-Nelson model (without any disorder).
Here, NHSE arises due to the non-reciprocal hopping parameter $\gamma$.
(b) $\lambda \ne 0$, $\alpha \ne 0$ and $\gamma\ne0$: the model reduces to a non-reciprocal NH AA model that includes a complex QP potential with the introduction of $\alpha$.
In the clean Hatano-Nelson model without an interface, all the bulk states accumulate at one of the edges (termed as NHSE), which is determined by the sign of the non-reciprocity parameter, $\gamma$.
The energy spectra for the periodic and open boundary conditions are markedly different, with the former creating closed loops in the complex plane that enclose the energy spectra corresponding to the open chain, which lies along the real axis \cite{Zhang}.
The reality of the OBC spectrum in the Hatano–Nelson model follows from the fact that its NH Hamiltonian can be mapped to a Hermitian Hamiltonian via a similarity transformation \cite{Okuma}.
Because similarity transformations leave the eigenvalues of a finite-dimensional matrix invariant, the OBC spectrum of the clean Hatano–Nelson Hamiltonian coincides exactly with that of its Hermitian counterpart.
Fig.~\ref{fig:1}(a) shows the energy spectrum of $H$ for $\lambda = 0$, which effectively represents a combination of a clean Hatano–Nelson model and its mirror image, joined back-to-back at the interface located at $k = L$.
As a result of this structure, the energy spectrum becomes entirely real in the absence of the QP potential.
The probability distribution of the corresponding eigenstates is shown in Fig.~\ref{fig:1}(b).
These eigenstates localize at exactly the interface ($k=L$).

In the presence of the QP potential ($\lambda\ne 0$, $\alpha\ne 0$) along with $\gamma\ne0$, the spectrum displays a hierarchical structure of three bands, as illustrated in Fig.~\ref{fig:1}(c).
In the case of $\gamma = \alpha = 0$, the system undergoes an AL transition at $\lambda = t$ \cite{QP}, governed by the self-duality of the model.
This property ensures the system is either fully extended or localized, dictated solely by the QP potential strength, $\lambda.$
The introduction of $\alpha$ modifies this behavior, shifting the AL transition to a critical value of $\alpha$, which is $\ln|t/\lambda|$.
This occurs because all the localized eigenstates in the self-dual space share a uniform inverse localization length of $\ln|\lambda/t|$ \cite{PhysRevLett.122.237601}.
With $\alpha=0$ but $\gamma\ne0$, the localization transition instead occurs at $\lambda = \text{max}(t+\gamma, t-\gamma)$ \cite{PhysRevB.100.054301}.
Using Avila's global theory \cite{10.1007/s11511-015-0128-7}, Li \textit{et al.} \cite{PhysRevB.110.134203} demonstrated that for a generalized scenario ($\alpha\ne0,\gamma\ne0$), the AL transition occurs at
\begin{equation}
    \alpha_c = \ln|\text{max}(t+\gamma, t-\gamma)/\lambda|
    \label{eq:alpha}
\end{equation}
Thus, Fig.~\ref{fig:1}(d) highlights this scenario through the probability distribution of eigenstates for a high value of $\alpha$ $(\alpha > \alpha_c)$, where the AL dominates the NHSE.
The above results are summarized in Table \ref{table:1}.
\begin{table}[h!]
\begin{tabular}{|c|c|c|}
\hline
NH parameters & $\alpha=0$ & $\alpha\ne0$ \\
\hline
$\gamma=0$ & $\lambda_c=t$ & $\alpha_c=\ln |t/\lambda|$ \\
\hline
\multirow{2}{3em}{$\;\gamma\ne0$} & $\lambda_c=$ & $\alpha_c=$\\
 & $\text{max}(t+\gamma,t-\gamma)$ & $\ln|\text{max}(t+\gamma,t-\gamma)/\lambda|$\\
\hline
\end{tabular}
\caption{The table presents the analytically determined delocalization-localization transition points as functions of the two NH parameters, namely, $\alpha$ (the imaginary phase of the QP potential) and $\gamma$ (the non-reciprocity parameter in the hopping amplitude $t$).}
\label{table:1}
\end{table}

Let us briefly recapitulate the origin and the role of the imaginary component ($i\alpha$) in the QP potential, namely, $\lambda_k = 2\lambda \cos(2\pi\beta k + i\alpha)$.
In a seminal work, Hatano and Nelson \cite{PhysRevLett.77.570} demonstrated that an imaginary vector potential induces a DL transition in a 1D randomly disordered system.
Following this, extensive theoretical and experimental efforts have explored the role of an imaginary vector potential, particularly in QP-disordered 1D systems.
Among those, the work of S. Longhi \cite{PhysRevLett.122.237601} stands out, where he showed that an imaginary gauge field, introduced via complex QP potential having parity-time ($\mathcal{PT}$) symmetry, the entire spectrum undergoes a DL transition along with $\mathcal{PT}$-symmetry-breaking when the imaginary gauge field exceeds a critical threshold value.
In our system, the underlying mechanism of the transition between AL and NHSE can be understood as follows.
The non-reciprocity comes into play in the form of $\gamma$, which effectively swaps its role with $\alpha$ under a discrete Fourier transformation.
Thus, when $\alpha$ exceeds a critical value $\alpha_c$, which corresponds to the decay length of skin modes arising from $\gamma$, a DL transition occurs, converting skin modes into Anderson-localized states.
In this way, $\alpha$ functions as a tunable imaginary gauge field that mediates the competition between boundary-localized NHSE states and disorder-induced Anderson localization.

\section{\label{s3}Time evolution of the NH AA model}
We have explored the localization properties of the eigenstates of the Hamiltonian for the NH AA model given by Eq.~\eqref{eq:Ham}.
An intriguing question arises: does the interplay between the NHSE and the AL persist over long time scales, or does the time-evolved system unveil any interesting physics?
To address this, we investigate the time evolution of an excited wavefunction in the NH AA model.
Let the initial wavefunction at $t=0$ be $\ket{\Psi(x, 0)}$, which can be expanded as a linear combination of the eigenstates of $H$ in Eq.~\eqref{eq:Ham}, given as
\begin{equation}
\ket{\Psi(x, 0)}=\sum_{q=1}^{2L+1}a_q(0)\psi_q(x),
\label{eq:a_q1}
\end{equation}
where $a_q(0)$ is the coefficient corresponding to the $q^{\text{th}}$ eigenstate ($\psi_q$) at $t=0$ and is responsible for the time evolution of the initial wavefunction, $\ket{\Psi(x, 0)}$.
Note that $x$ is a discrete variable and lies in the range $x\in[1,2L+1].$
As, the eigenstates, $\psi_q(x)$ of $H$ form a complete orthonormal basis, we can express $a_q(0)$ and $a_q(t)$ as,
\begin{align}
    a_q(0)&=\braket{\psi_q(x)|\Psi(x, 0)}\nonumber\\
    \Rightarrow a_q(t)&=\braket{\psi_q(x)|\Psi(x, 0)}e^{-\frac{iE_qt}{\hbar}},
    \label{eq:a_q2}
\end{align}
for any $q$, where $E_q$ is the eigenvalue corresponding to the eigenvector, $\psi_q(x)$.

\begin{figure}
    \centering
    \includegraphics[width=\linewidth]{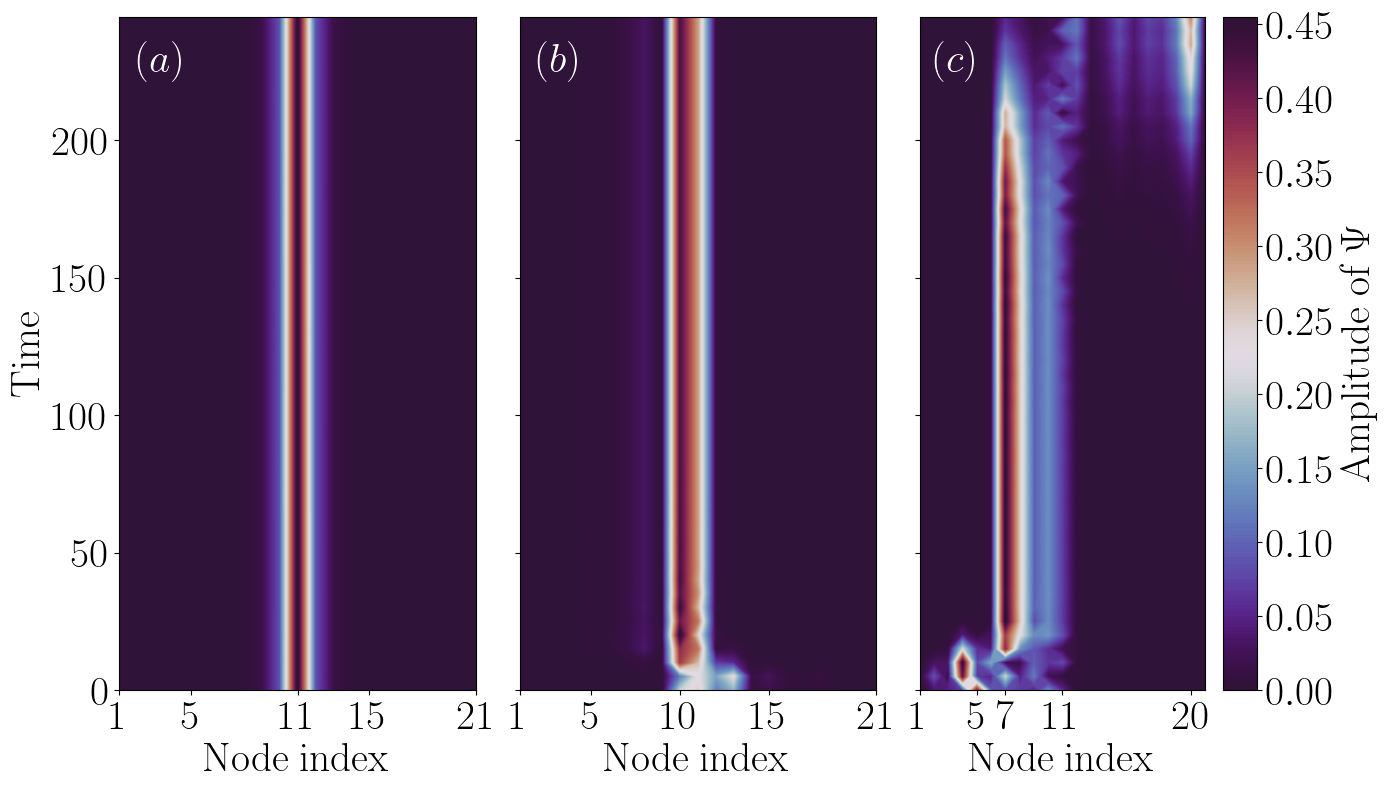}
    \caption{The temporal evolution of the excitation in our model is illustrated for three scenarios: (a) $\lambda = 0$, (b) $\lambda = 1,\; \alpha = 0.425$, and (c) $\lambda = 1,\; \alpha = 0.9$. In the first case, the localization of the wavefunction at the interface is attributed to the NHSE. In the second case, at $\alpha = \alpha_c$, a slight shift in the localization is observed. Finally, for a much larger value of $\alpha$, the wavefunction undergoes brief transitions, namely, from the $5^{\text{th}}$ site to the $4^{\text{th}}$ site, then to the $7^{\text{th}}$ site for an extended period, and ultimately localizes at the $20^{\text{th}}$ site.}
    \label{fig:2}
\end{figure}
Now, as an initial condition, we choose a delta-type excitation of the form,
\begin{equation}
    \ket{\Psi(x, 0)}=\delta(x-m)
    \label{eq:delta}
\end{equation}
which is localized entirely at the $m^{\text{th}}$ site and is zero elsewhere.
The time evolution of $\ket{\Psi(x, 0)}$ from Eqs.~\eqref{eq:a_q1} and \eqref{eq:a_q2}, can be expressed as,
\begin{equation}
    \ket{\Psi(x, t)}=\sum_{q=1}^{2L+1}\braket{\psi_q(x)|\Psi(x, 0)}e^{-\frac{iE_qt}{\hbar}}\psi_q(x).
    \label{eq:psi}
\end{equation}
However, this equation applies explicitly to systems without boundaries, where $ x $ ranges from $ -\infty $ to $ +\infty$.
Hence, assuming that the wavefunction yields vanishing probability density at the edges of the chain, any reflection of the state during its evolution is thereby precluded.
Subsequently, we can reliably use Eq.~\eqref{eq:psi} to determine $\ket{\Psi(x, t)}$ for any finite range of $x$ at all subsequent times.
\begin{figure*}[t]
    \centering
    \includegraphics[trim={0 2cm 0 4cm},clip,width=\linewidth]{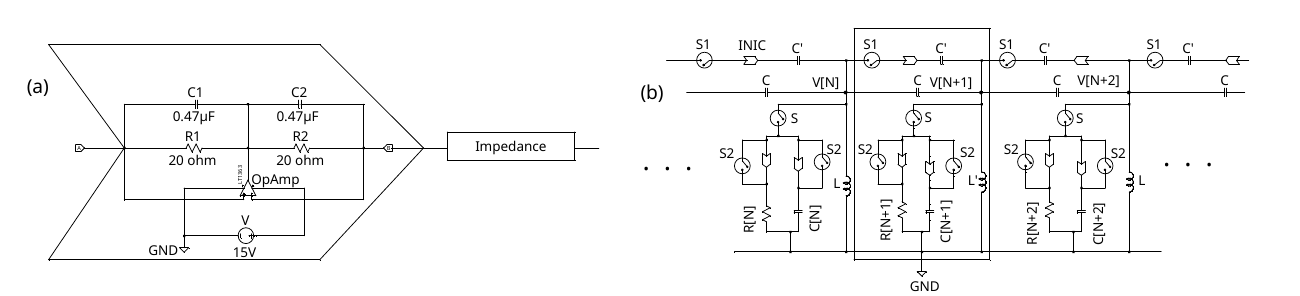}
    \caption{(a) Circuit diagram of an INIC is shown. The current entering the INIC from the left(right) side experiences a negative(positive) impedance placed in the rectangular box. (b) The TEC diagram corresponds to the non-reciprocal NH AA model. The rectangular box highlights the $(N+1)^{\text{th}}$ node. $V[N]$ denotes the output voltage at $N^{\text{th}}$ node.}
    \label{fig:3}
\end{figure*}

NH systems are known for their non-conservation of energy, leading to non-unitary time evolution.
As a result, these systems violate probability conservation, resulting in the norm of the wavefunction either growing or diminishing as a function of time.
Thus, we have to normalize the amplitude of the evolved wavefunction with its norm at each small interval $ dt $, which occurs through a two-step process \cite{PhysRevB.107.064305}.
First, the wavefunction evolves as,
\begin{equation}
    \ket{\Psi(x,t+dt)}=e^{-\frac{iHdt}{\hbar}}\ket{\Psi(x,t)}.
    \label{eq:psi1}
\end{equation}  
This is hence followed by a normalization step,
\begin{equation}
    \ket{\Psi(x,t+dt)}=\frac{\ket{\Psi(x,t+dt)}}{||\ket{\Psi(x,t+dt)}||},
    \label{eq:psi2}
\end{equation}  
where $ ||.|| $ denotes the norm.

Let us now employ Eqs.~\eqref{eq:psi1} and \eqref{eq:psi2} to numerically analyze $\ket{\Psi(x, t)}$.
As an example, we excite the delta-type wavefunction at the $5^{\text{th}}$ site, that is, $m=5$ in Eq.~\eqref{eq:delta}.
Fig.~\ref{fig:2}(a) illustrates the NHSE observed at the interface in the absence of any QP potential $(\lambda = 0).$
Fig.~\ref{fig:2}(b) represents the system at a critical value of $\alpha$, given by $\alpha_c = \ln|\text{max}(t+\gamma, t-\gamma)/\lambda| \simeq 0.425$ for $\lambda = 1$.
At this point, the eigenstates of $H$ undergo a transition from exhibiting NHSE to AL, and the localization of the time-evolved wavefunction starts to shift away from the interface.
For $\alpha > \alpha_c$, $\ket{\Psi(x, t)}$ no longer remains localized at the interface, as shown in Fig.~\ref{fig:2}(c).
Instead, the wave propagates via quantized jumps between the AL states located randomly at distinct sites of the chain, a phenomenon termed as `NH jumps' \cite{Weidemann2021, PhysRevResearch.3.013208, PhysRevB.106.064205}.
These jumps are distinctive artefacts of the NH disorder, incorporated via the parameter $\alpha$ in our case.
Thus, the time evolution of the wavefunction involves several NH jumps over a specific time frame, as depicted in Fig.~\ref{fig:2}(c).
The jumps can be predicted with the help of both the initial wavefunction $\ket{\Psi(x,0)}$ and $a_q(t)$.
Note that $a_q(t)$ depends on the corresponding eigenvalues, given by Eq.~\eqref{eq:a_q2}.

\section{\label{s4}TEC construction}
Similar to the Hamiltonian of a TB model, electrical circuit networks operate based on their Laplacians, which govern the network's response at each node \cite{F_Y_Wu_2004}.
For an electrical network with $N_0$ nodes, let $\mathcal{L}$ represent the Laplacian, and $V_i$ and $I_i$ denote the voltage and the total current through an external source at the $i^{\text{th}}$ node.
According to Kirchhoff's law, the following relation holds,
\begin{equation}
I_i=\sum_{p(i\ne p)}^{N_0}X_{ip}(V_i-V_p)+X_iV_i\quad\text{for}\quad i=1, 2, 3, \ldots, N_0,
\label{eq:current}
\end{equation}  
where $X_{ip}$ is the conductance between $i^{\text{th}}$ node and $p^{\text{th}}$ node.
Note that $X_{ii}$ has no physical meaning and is set to zero, while $X_i$ represents the resultant conductance between $i^{\text{th}}$ node and the ground.
With these definitions, Eq.~\eqref{eq:current} can be expressed as $I=\mathcal{L}V$, where $\mathcal{L}$ is the $N_0\times N_0$ Laplacian matrix with elements, $\mathcal{L}_{ip}=-X_{ip}+\delta_{ip}W_i,$ where $W_i=\sum_{p}X_{ip}+X_i.$
Thus, the Laplacian $\mathcal{L}$ effectively replicates a specific second-quantized Hamiltonian of a TB model, owing to the direct resemblance between Kirchhoff's current laws and the structure of the TB model.

\subsection{TEC: NH AA model}
Now, we focus on forming an analog circuit corresponding to the TB model given by Eq.~\eqref{eq:Ham}.
To achieve the goal, the Laplacian of the circuit must accurately replicate the Hamiltonian at the resonant frequency, $f_R$, of the circuit.
The inter-site hoppings can be modeled by capacitors $(C)$, while the non-reciprocity in the hoppings $(\gamma)$ is introduced via negative impedance converter (INIC) in the circuit \cite{Helbig2020}.
As shown in Fig.~\ref{fig:3}(a), for the realization of the INIC, an operational amplifier (opamp) is employed.
An opamp is a differential amplifier, that is, its output voltage is proportional to the difference of the positive and negative input potential, with the amplification being much larger than unity.
The opamp is operated in negative feedback configuration, which means that its output is coupled to its inverting (negative) input via an impedance ($C1$ and $C2$ in our case).
To ensure the stability of the circuit, two resistors ($R1=R2=20\,\Omega$) are placed in parallel to the capacitors, $C1$ and $C2$, as shown in Fig.~\ref{fig:3}(a).
The resonant frequency of the circuit is given by,
\begin{equation}
    f_R=\frac{1}{2\pi\sqrt{2LC}}\simeq 5191\,Hz;\quad\omega_R=\frac{1}{\sqrt{2LC}},
    \label{eq:freq}
\end{equation}
with $L=10\,\mu$H and $C=47\,\mu$F.
All these values are standard for commercial uses and are kept fixed throughout this work.
The real and the imaginary parts of the QP potential, $\lambda_k$ in Eq.~\eqref{eq:Ham}, are represented by node-dependent capacitors $(C[k])$ and resistors $(R[k])$, respectively, with $|\text{Re}(\lambda_k)|\equiv\omega_R\, C[k]$, and $|\text{Im}(\lambda_k)|\equiv [R[k]]^{-1}$.
Here, $k$ denotes the node index and is analogous to the site index of Eq.~\eqref{eq:Ham}.
The detailed mathematical rigor behind the formation of the Laplacian of this TEC is provided in Appendix~\ref{a1}.
The switches for the circuit elements are denoted by $S$ and $S1$.
The circuit also includes master switches for $ S $ and $ S1 $ (not shown in Fig.~\ref{fig:3}(b)), which control all the $ S $ and $ S1 $ switches across the circuit.
For instance, the master switch for $ S $ ($S1$) can simultaneously open or close all $ S $ ($S1$) switches.
However, this functionality does not extend to $ S2 $, as $ S2 $ is specifically designed to alter the signs of $ C[k] $ or $ R[k] $ based on the values of $ \lambda_k $, which, in turn, depend on the node index, $k$.
The total number of nodes is $(2N+1)=21$, which is kept fixed throughout the analysis.
It is worth noting that $21$ nodes are sufficient to obtain reliable results from the TEC framework using LTspice software \cite{LTspice}, which provides realistic results that align closely with experimental observations.
Let us briefly summarize the measurement process and the data acquisition thereafter using LTspice.
Users can select from LTspice’s built-in device models or define their own.
To construct a circuit in the software, elements from its library (or customized models) are placed on the schematic to obtain a desired circuit diagram, followed by appropriate connections.
Once assembled, the circuit is simulated to observe its response under various conditions, including the presence or absence of external sources.
It is important to note that LTspice presents the node voltages and currents through circuit elements as functions of time using a graphical interface.

To get results in support of the theoretical results in Fig.~\ref{fig:1}, we have to obtain the eigenvectors of the Laplacian in terms of measurable quantities like the voltage or the impedance profile.
However, to do that, every node must be excited via a current (or a voltage) source, making the process unnecessarily complicated.
Instead, a more practical approach is to excite a single node using a current (or a voltage) source and simulate the voltage response of the TEC using the LTspice software.
The calculations still allow us to observe the localization of the VP, which is equivalent to NHSE and AL in the TB model (Figs.~\ref{fig:1}(b) and \ref{fig:1}(d)).
Interestingly, both these phenomena are tunable in our TEC.
The detailed analysis of the time evolution of the excitation is explained in Appendix~\ref{a2}.

\subsection{NHSE in TEC}
To realize and explore the NHSE in the non-reciprocal circuit, we open (disconnect) the master switch for all the $ S $ switches and close (connect) the master switch for all the $ S1 $ switches in Fig.~\ref{fig:3}(b), thereby configuring the TEC to replicate the clean Hatano-Nelson model with an interface.
To demonstrate a voltage build-up, we excite the $3^{\text{rd}}$ node with a voltage pulse of amplitude $1$ mV for a duration of $ 10 \, \mu$s and record the output signal at each node for $3000\,\mu$s.
To effectively analyze the output, we calculate the root mean square (RMS) values of the voltage signal at suitable time intervals for each node.
It is important to note that we have verified that all the relevant phenomena are captured with this time range ($3000\,\mu$s).
Thus, for this particular setup, the observations are made until $t=3000\,\mu$s, corresponding to the maximum value along the $y$-axis (representing time).

\begin{figure}
    \centering
    \includegraphics[width=\linewidth]{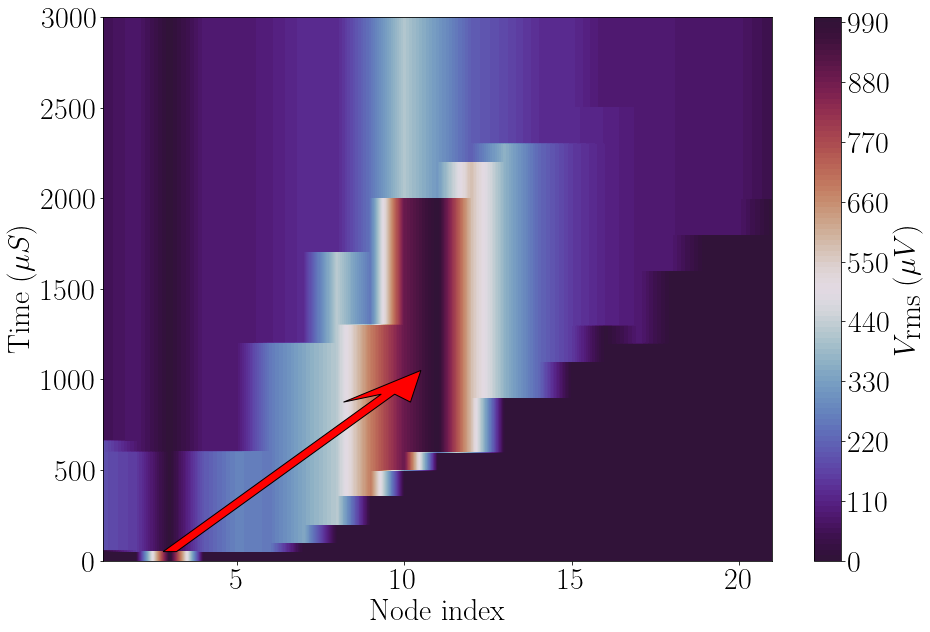}
    \caption{Colormap of the RMS values of the output VP as functions of both time and node index is shown. Commencing around time equal to $500\,\mu$s and continuing until approximately $2000\,\mu$s, the VP becomes localized at the interface, whose location occurs at the $11^{\text{th}}$ node. The red arrow illustrates the progression of the excitation from the $3^{\text{rd}}$ node to the interface over time.}
    \label{fig:4}
\end{figure}

Using LTspice, these data are visualized in a colormap representing the VP as a function of time in Fig.~\ref{fig:4}.
The results show that the VP localizes at the interface, specifically at $11^{\text{th}}$ node, as illustrated in Fig.~\ref{fig:4}.
This behavior closely reflects the NHSE observed in our TB model (Fig.~\ref{fig:1}(b)).
Moreover, this interface localization of the VP serves as a direct representation of the time-evolved wavefunction in the TB model (see Fig.~\ref{fig:2}(a)), where the delta excitation at the $5^{\text{th}}$ site ultimately localizes at the $11^{\text{th}}$ site.
The grounded inductors at the edges have a distinct value of $ L_{\text{edge}} \simeq 62 \, \mu$H, while the inductor at the $ 11^{\text{th}} $ node, representing the interface, is chosen as $ L' \simeq 6 \, \mu$H.
These values differ from the other grounded inductors, $L(=10\,\mu\text{H})$.
To ensure that the resonance conditions required by the Laplacian of the TEC (see Eq.~\eqref{eq:rca} of Appendix~\ref{a1}) are satisfied for faithfully reproducing the TB model, we choose specific values for $L_{\text{edge}}$ and $L'$.
The role of the non-reciprocity parameter, $ \gamma $, is implemented using $ C'\;(=32\,\mu\text{F})$ along with an INIC (see Fig.~\ref{fig:3}(b)).
The mapping between the TB model and the circuit is thus established via the relations $ (t \pm \gamma) \equiv \omega_R (C \pm C') \simeq (1.53 \pm 1.04) $.
Recently, Liu \textit{et al.} \cite{PhysRevResearch.5.043034} demonstrated NHSE at the interface in TECs for both 1D and 2D systems by employing a voltage follower, where the current flows unidirectionally, instead of using an INIC.
Both approaches are well-accepted and provide reliable results for constructing non-reciprocal circuits.

Let us now illustrate the measurement procedure to generate Fig.~\ref{fig:4} and its production from the raw data obtained from LTspice.
Since the excitation (voltage pulse) at the $3^{\text{rd}}$ node is short-lived and vanishes after $10\,\mu$s, localization for a short duration appears at the $3^{\text{rd}}$ node in Fig.~\ref{fig:4} when the voltage is measured.
Consequently, the circuit’s response is temporally constrained, implying that the output voltage at any node decays with time due to the presence of dissipative elements, such as the resistors (embedded within the INIC).
This makes smooth voltage measurement as a function of time (at very short intervals) challenging.
Therefore, the most effective approach is to measure the RMS values of the output over non-equivalent discrete time intervals.  
For instance, the output voltage at the $11^{\text{th}}$ node (interface) is measured at the following time intervals:
(i) $0-600\,\mu$s,\quad
(ii) $600-2000\,\mu$s, and\quad  
(iii) $2000-3000\,\mu$s.
These chosen time intervals vary from node to node, depending on where the output signal reaches its maximum amplitude and how long it is sustained before decaying.
This measurement process is responsible for the appearance and disappearance of the voltage localization occurring in a stepwise manner, observed in Fig.~\ref{fig:4}.
Furthermore, the data along the $z$-axis, representing the RMS values of output voltages at different nodes and time intervals, are linearly interpolated using Python programming language to ensure smooth transitions while preserving the integrity of the raw data obtained from LTspice.

Interestingly, no signal is detected at the $11^{\text{th}}$ node up to $600\,\mu$s, as the RMS value of the output voltage in this interval is nearly zero.
However, during the $600-2000\,\mu$s interval, the RMS value increases significantly, leading to the `dark' region (approximately $990\,\mu$V) in Fig.~\ref{fig:4}.
Alternatively, this can be understood by noting that, initially excited at the $3^{\text{rd}}$ node, the signal requires $600\,\mu$s to reach the interface at the $11^{\text{th}}$ node, represented by the red arrow, and hence, no output signal is observed at the $11^{\text{th}}$ node before $t=600\,\mu$s.
Similarly, in the $2000-3000\,\mu$s interval, the RMS output voltage diminishes due to dissipation in the circuit.
This localization occurs regardless of the node of excitation and the pulse amplitude of the pulse (or the form of the pulse, such as square, triangular, etc.), demonstrating the robustness of the phenomenon.
Interestingly, it parallels the phenomenon of `topological funneling of light,' where a light field within a photonic mesh lattice with an interface is directed toward the interface, irrespective of their shape or the input location \cite{Weidemann2020}.

\subsection{AL in TEC}
As depicted in Fig.~\ref{fig:3}(b), to isolate and observe the AL, we close the master switch for $S$ while keeping $ S1 $ open.
The only practical technique to incorporate the onsite QP potential in the circuit is to place the capacitors and resistors obeying the following equations, namely,
\begin{align}
    C[k]&=-\text{Re}(\lambda_k)/\omega_R=-2\lambda\cos {(2\pi\beta k)}\;\cosh{\alpha}/\omega_R,\\
    R[k]&=[\text{Im}(\lambda_k)]^{-1}=\left[-2\lambda\sin {(2\pi\beta k)}\;\sinh{\alpha}\right]^{-1},
    \label{eq:rc}
\end{align}
\begin{figure}[t]
    \centering
    \includegraphics[width=\linewidth]{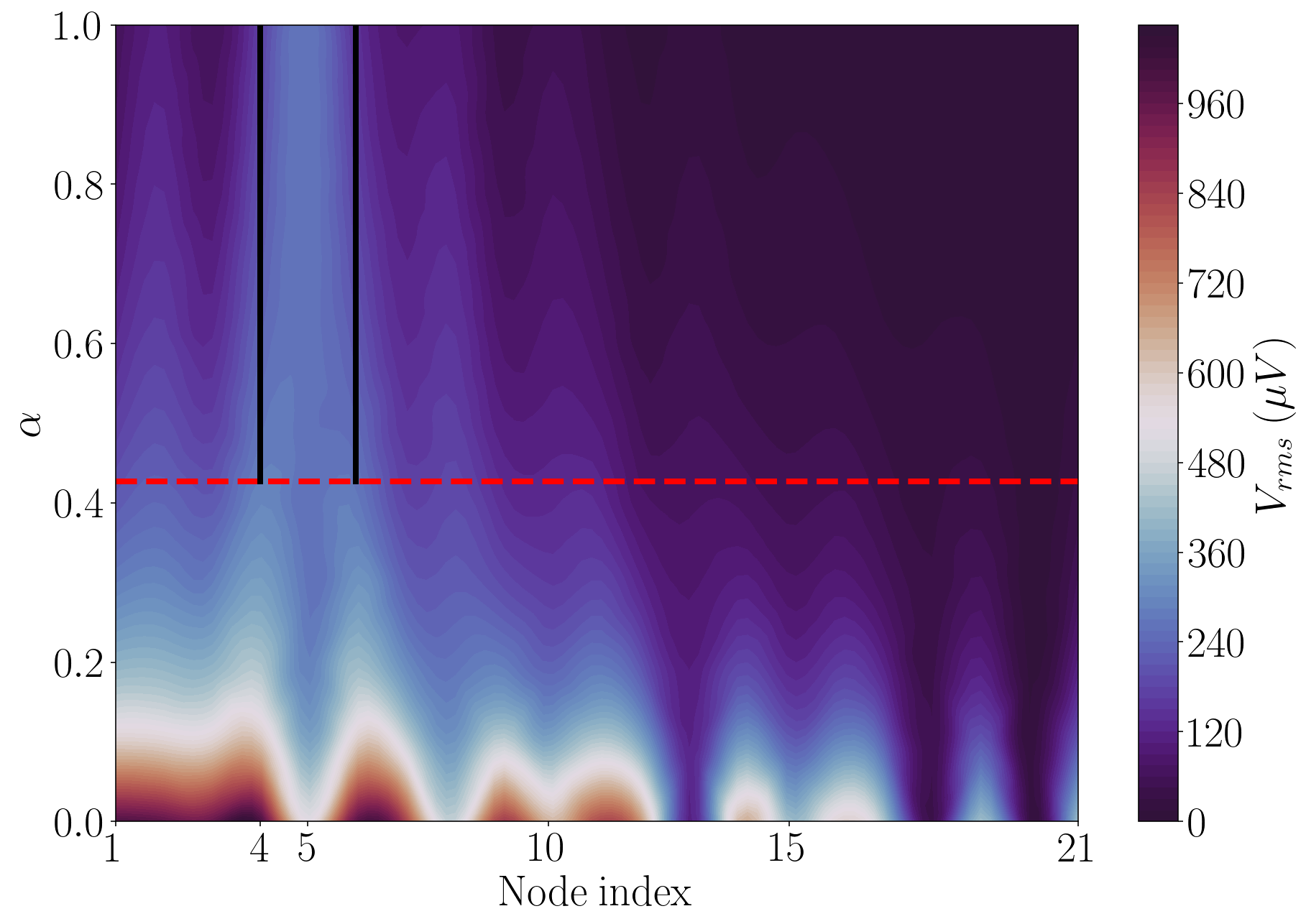}
    \caption{The RMS values of the output signal, measured over the duration from $600\,\mu$s to $2000\,\mu$s, are presented as functions of the node indices and $\alpha$. The red dashed line denotes the critical value of $\alpha$, namely, $\alpha_c$, obtained from Eq.~\eqref{eq:alpha}. The two black vertical lines indicate the localization of the VP at the $5^{\text{th}}$ node for $\alpha>\alpha_c.$}
    \label{fig:5}
\end{figure}
derived from the expression for $\lambda_k$ in Eq.~\eqref{eq:Ham} with $\beta^{-1}$ being the golden ratio.
It is crucial to note that both $ \text{Re}(\lambda_k) $ and $ \text{Im}(\lambda_k) $ being oscillatory functions, can assume negative values, and consequently, $ C[k] $ and $ R[k] $ may be negative.
To avoid negative capacitance values, a capacitor of constant value can be grounded \cite{PhysRevB.108.144203}, ensuring no impact on the central results.
However, this strategy is not feasible for resistors, as additional grounded resistors would unnecessarily increase dissipation in the circuit.
As a remedy, we implement negative resistors using INICs.
While the primary role of the INIC is to introduce non-reciprocal current flow between nodes, its impact on the circuit's Laplacian becomes negligible when one of the connected nodes is grounded, as is the case in our setup.
This can be understood as follows.
An INIC, placed in series with an impedance, introduces a positive impedance in one direction and a negative impedance of equal magnitude in the opposite direction, thereby breaking reciprocity.
When placed in series with an impedance that connects a node to ground, the impedance seen from that particular node becomes negative.
However, relative to the ground, the impedance remains positive.
Since the Laplacian is constructed by applying Kirchhoff’s current law at each node (excluding the ground), only the impedance as viewed from the node is relevant.
Therefore, in spite of breaking the reciprocity, the use of INICs still allows us to realize negative impedance in an anticipated manner, eliminating the non-reciprocal effects.
To get $-R[k]$ or $-C[k]$, the absolute values of them are placed at each node based on the values of $ |\text{Re}(\lambda_k)| $ and $ |\text{Im}(\lambda_k)| $, respectively, and $ S2 $ switches are carefully toggled (closed or opened) at each node to fix the signs of $\text{Re}(\lambda_k)$ and $\text{Im}(\lambda_k)$.
Moreover, the grounded inductors at the edges have a value of $L_{\text{edge}}=20 \, \mu$H, while $L'$ is identical to $L$ as there exists no interface in this case.

Fig.~\ref{fig:5} illustrates the VP of this TEC setup, analogous to the reciprocal NH AA model, as a function of $\alpha$, with $\lambda = 1$ in Eq.~\eqref{eq:Ham}.
A constant current source of amplitude $1$ mA and frequency $f_R$, as defined in Eq.~\eqref{eq:freq}, is applied at the $4^{\text{th}}$ node.
This leads to a smooth variation of VP, in contrast to the stepwise pattern observed in Fig.~\ref{fig:4}.
Consequently, there is no need to compute the RMS value of the output voltage over different time segments for different nodes.
For the reciprocal NH AA model, the AL transition should occur at the critical value $\alpha_c=\ln|t/\lambda|$, calculated using Eq.~\eqref{eq:alpha}, with $\gamma = 0.$
The red dashed line in the figure marks this critical value, $\alpha_c \simeq 0.425$.
However, a sharp transition is not observed due to practical factors such as the finite system size (only $21$ nodes being considered) and the simulational limitations of LTspice.
Nevertheless, the results demonstrate that for $\alpha > \alpha_c$, the VP becomes predominantly localized at the $5^{\text{th}}$ node.
When our simulation is repeated by considering excitation at different nodes (not shown here), the localization consistently occurs in the vicinity of the respective excitation node, thus highlighting a predictable and robust localization center.
The predictability sharply contrasts with the NH jumps observed in our TB model in Fig.~\ref{fig:2}(c), where wave packet evolution transpires not through gradual diffusion but sudden transitions between the distinct states.
This discrepancy arises from the nature of the input excitation, as the time evolution of the VP is highly sensitive to the form of the initial stimulus.
While the NH jumps in the NH AA model result from the spontaneous evolution of an initially localized delta-type wavefunction, the VP localization for the TEC in Fig.~\ref{fig:5} emerges under a steady sinusoidal current source, thoroughly discussed in Appendix~\ref{a2}.

\subsection{Competition between AL \& NHSE in TEC}
\begin{figure}[h]
    \centering
    \includegraphics[width=\linewidth]{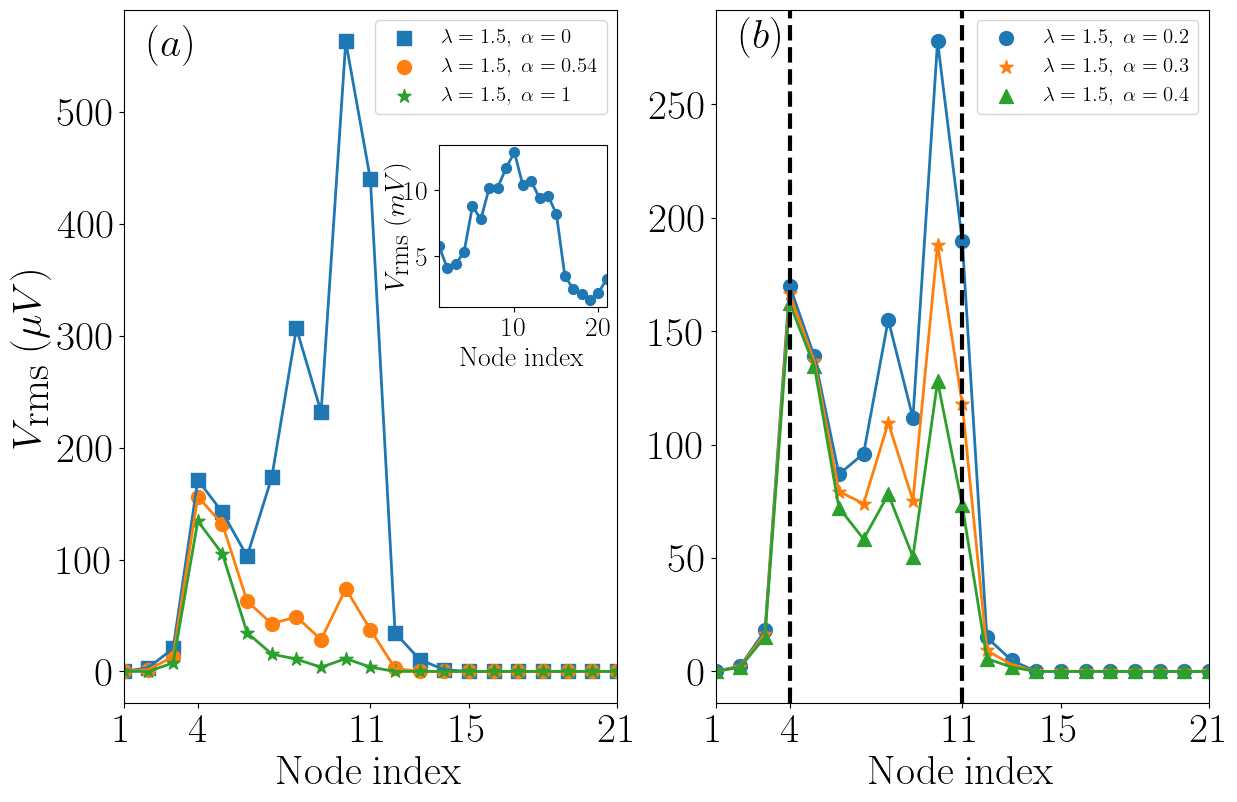}
    \caption{(a) The RMS values of the VP are plotted for different values of $\alpha$ at $\lambda = 1.5$. The time interval $1000\,\mu\text{s}-3000\,\mu\text{s}$ is used as a prototype over which the integration is carried out to calculate the RMS voltage for improved accuracy. The VP moves towards the $4^{\text{th}}$ node at $\alpha=\alpha_c$ with a much smaller amplitude. The inset depicts the VP without any QP potential $\lambda=0$, measured over the duration $2000\,\mu$s to $5000\,\mu$s. Note that all the values are in mV, which suggests voltage amplification, compared to the constant external signal, for a sufficiently longer time. (b) The same profile, but for values of $\alpha$ between zero and the critical value $(\alpha_c).$ The two black-dashed vertical lines enclose the key nodes (exciting and interface nodes) that contain the partial delocalization of the VP.}
    \label{fig:6}
\end{figure}

The interplay between NHSE and AL in the NH AA model exhibits fascinating behavior in the localization transition.
AL directs a single-site excitation toward a `{\it focal point},’ determined by the weight factors based on the overlap between the initial excitation and exponentially localized eigenstates \cite{Weidemann2021}, while NHSE drives it to an interface (or the edges).
To probe deeper into this interplay in the TEC, we close both $ S $ and $ S1 $ to incorporate both the non-reciprocity and the QP potential and again excite the $4^{\text{th}}$ node with a current source, as described earlier.
Fig.~\ref{fig:6}(a) shows that when the QP disorder is real $ (\alpha = 0) $, the disorder potential is weak compared to the non-reciprocity parameter $(\gamma)$ and is unable to drive the signal towards the excitation node.
However, once $ \alpha $ surpasses the critical value $ \alpha_c \simeq 0.54 $ (determined using Eq.~\eqref{eq:alpha}), the localization of the output voltage shifts to the excitation ($4^{\text{th}}$) node.
Remarkably, in contrast to Fig.~\ref{fig:2}(c) for the TB model, no NH jumps are observed in the TEC.
This distinction arises from the nature of the input current $I(t)$, which is a continuous sinusoidal signal in the circuit, as opposed to a delta-type excitation used in the TB model (given by Eq.~\eqref{eq:delta}).
As a result, the VP stabilizes in the vicinity of the excitation node (see Appendix~\ref{a2} for more detail).

The VP for $\alpha = 1$ represents the analogous scenario in the TEC corresponding to the non-reciprocal NH AA model depicted in Fig.~\ref{fig:1}(d).
An intriguing aspect of this behavior is that the output amplitude diminishes with increasing $\alpha$.
This behavior is attributed to the significant rise in $ |R[k]| $, which scales as $ [\sinh{\alpha}]^{-1} $ (Eq.~\eqref{eq:rc}).
Consequently, by tuning $ \alpha $, one can effectively manipulate both the spatial localization and the amplitude of the output signal.
This dual ability to modulate the NHSE and the AL dynamics is a distinctive feature of our TEC with potential applications similar to information transfer communication devices or developing highly sensitive sensors.
Upon closer inspection, a gradual increase in $ \alpha $, from zero to $ \alpha_c $ while keeping $ \lambda $ fixed ($ \lambda=1.5 $ in this case) unveils a peculiar phenomenon.

As previously discussed, AL is spatially confined over a short range near the excitation node in the TEC.
On the other hand, NHSE drives the excitation towards the interface.
This interplay generates a fascinating \textit{tug-of-war} scenario between the NHSE and the AL-induced localization, resulting in a `partial' delocalization of the output signal within a particular spatial range.
This range is bounded by two controllable key nodes, namely, the interface ($11^{\text{th}}$) and the excitation ($4^{\text{th}}$) node.
Fig.~\ref{fig:6}(b) illustrates this phenomenon, where the VP shows non-zero oscillations between the $4^{\text{th}}$ and the $11^{\text{th}}$ nodes, beyond which VP decays to zero.
During the transition, the amplitude of the output signal becomes somewhat uniform across the intermediate nodes, effectively creating a spatial channel for the signal.
Furthermore, the position and width of this channel can be tuned by altering the excitation node or the interface of the circuit, offering versatile control over the behavior of the input signal to propagate.
Thus, as said earlier, $ \alpha $ serves as a critical parameter, acting as a switch that toggles between the phenomenon of NHSE and AL on the spreading dynamics of a single-site excitation.

\section{\label{s5}Conclusion}
\noindent In this work, we have investigated the interplay between the non-Hermitian skin effect and Anderson localization in a one-dimensional chain, where non-reciprocal hopping amplitudes drive the former, while quasiperiodic disorder induces the latter.
Using a non-Hermitian variant of the Aubry-Andr\'e model, we explored the localization properties and analyzed its time evolution under single-site excitation.
Additionally, we have analyzed the time evolution to reveal the occurrence of quantum jumps between skin states and Anderson-localized states within the system.
To bridge the gap between theory and experiment, we have proposed a topolectrical circuit to realize the model.
Specifically, we observe a distinct behavior in circuits, where the excitation voltage remains localized near the excitation node, in contrast to exhibiting non-Hermitian jumps, as predicted in the model.
Our designed circuit possesses the ability to regulate the voltage profile across the circuit network, specifically between the excitation node and the interface, which also leads to partial delocalization of the output voltage.
The competition between Anderson localization and skin effect manifests in a tangible and experimentally accessible manner, gaining precise control over the output and establishing electrical circuits as a powerful platform for studying such effects.
This quantum-to-classical correspondence not only uses the principles rooted in quantum localization phenomena but also enhances our understanding of them, and paves the way for designing advanced sensors and devices for communication and information by leveraging the tunability and versatility of topolectrical circuits.

\section*{DATA AVAILABILITY}

The data that support the findings of this article are openly available in Ref.~\cite{codes}.

\appendix

\section{\label{a1}Laplacian of the TEC}
The TEC corresponding to the Hamiltonian, $H$ (Eq.~\eqref{eq:Ham}), is constructed and analyzed in section~\ref{s4}A.
For an electrical network with $2N+1$ nodes, which is $21$ for this case $(N=10)$, let $\mathcal{L}$ represent the Laplacian, and $V_i$ and $I_i$, respectively, denote the voltage and the total current from an external source at $i^{\text{th}}$ node.
Let us now derive the Laplacian of the circuit (refer to Fig.~\ref{fig:3}(b)) under the condition where the master switch for $S$ is open and $S_1$ is closed, representing the non-reciprocal circuit without the QP potential.
Following Eq.~\eqref{eq:current} without any external source and the following $2N+1$ equations,
\onecolumngrid
\begin{align*}
    \frac{1}{j\omega L_{\text{edge}}}V_1+j\omega(C-C')(V_1-V_2)&=0,\\
    \frac{1}{j\omega L}V_2+j\omega(C+C')(V_2-V_1)+j\omega(C-C')(V_2-V_3)&=0,\\\vdots\\
    \frac{1}{j\omega L'}V_{N+1}+j\omega(C+C')(V_{N+1}-V_{N})+j\omega(C-C')(V_{N+1}-V_{N+2})&=0,
\end{align*}
\begin{align*}
    \vdots\\
    \frac{1}{j\omega L}V_{2N}+j\omega(C-C')(V_{2N}-V_{2N-1})+j\omega(C+C')(V_{2N}-V_{2N+1})&=0,\\
    \frac{1}{j\omega L_{\text{edge}}}V_L+j\omega(C-C')(V_{2N+1}-V_{2N})&=0,
\end{align*}
where $V_i$ is the voltage at $i^{\text{th}}$ node, and the values of the circuit elements, $C,\, C',\, L,\, L'$ and $L_{\text{edge}}$ are defined in the sections~\ref{s4}.A and \ref{s4}.B.
Note that $j$ represents the imaginary number $(=\sqrt{-1})$ and $(N+1)^{\text{th}}$ node denotes the interface.
When we formulate these $2N+1\;(=21)$ equations corresponding to Kirchhoff's law in the form $ I = \mathcal{L}V $, the resulting Laplacian matrix, $\mathcal{L}$ is given as,
\begin{equation}
    \mathcal{L}(\omega) = \begin{pmatrix}
    \frac{1}{j\omega L_{\text{edge}}}+j\omega(C-C') & -j\omega(C-C') & 0 & \cdots\\
    -j\omega(C+C') & \frac{1}{j\omega L}+2j\omega C & -j\omega(C-C') & 0 & \cdots \\
    \vdots & \vdots & \vdots & \vdots & \vdots\\
    \cdots 0 & -j\omega(C+C') & \frac{1}{j\omega L'}+2j\omega (C+C') & -j\omega(C+C') & 0 \cdots \\
    \vdots & \vdots & \vdots & \vdots & \vdots\\
    \cdots & 0 & -j\omega(C-C') & \frac{1}{j\omega L}+2j\omega C & -j\omega(C+C')\\
     & \cdots & 0 & -j\omega(C-C') & \frac{1}{j\omega L_{\text{edge}}}+j\omega(C-C')
    \end{pmatrix}
\end{equation}
Note that the diagonal terms vanish at the resonant frequency,
\begin{equation}
    \omega_R=\frac{1}{\sqrt{2LC}}=\frac{1}{\sqrt{L_{\text{edge}}(C-C')}}=\frac{1}{\sqrt{2L'(C+C')}}.
    \label{eq:rca}
\end{equation}
Subsequently, at $\omega=\omega_R$, $\mathcal{L}$ becomes,
\begin{equation}
    \mathcal{L}(\omega_R) = -j\begin{pmatrix}
    0 & \omega_R(C-C') & 0 & \cdots\\
    \omega_R(C+C') & 0 &  \omega_R(C-C') & 0 & \cdots \\
    \vdots & \vdots & \vdots & \vdots & \vdots\\
    \cdots 0 & \omega_R(C+C') & 0 & \omega_R(C+C') & 0 \cdots \\
    \vdots & \vdots & \vdots & \vdots & \vdots\\
    \cdots & 0 & \omega_R(C-C') & 0 & \omega_R(C+C')\\
     & \cdots & 0 & \omega_R(C-C') & 0
    \end{pmatrix}\label{eq:L1}
\end{equation}
\twocolumngrid
Thus, the Laplacian, $\mathcal{L}$, effectively replicates the matrix form of $H$ with $\lambda=0$, including a scaling factor of $-j$.
Consequently, the eigenvalues and the eigenvectors of $\mathcal{L}$ directly correspond to those of $H$ with $\lambda=0.$
This equivalence ensures that the localization properties of the eigenvectors of the TB model are reflected in the VP derived from the Laplacian.
Note that the capacitors are analogous to the hopping terms in the TB model, where $(t \pm \gamma) \equiv \omega_R(C \pm C')$.

Let us incorporate the QP potential, as described in Eq.~\eqref{eq:Ham}, into the TEC.
This addition introduces extra diagonal terms in $\mathcal{L}$, stemming from the node-specific values of the capacitors, $ C[k] $, and the resistors, $ R[k] $ at the $k^{\text{th}}$ node.
According to Eq.~\eqref{eq:Ham}, the QP potential is expressed as,
\begin{equation*}
    \lambda_k=2\lambda\left[\cos {(2\pi\beta k)}\cosh {\alpha}-i\sin {(2\pi\beta k)}\sinh {\alpha}\right].
\end{equation*}
Thus, at the resonant frequency, $\omega_R$, $C[k]$ and $R[k]$ assume following forms,
\begin{widetext}
    \begin{align*}
    C[k]=-\text{Re}(\lambda_k)/\omega_R=-2\lambda\cos {(2\pi\beta k)}\;\cosh{\alpha}/\omega_R,\quad
    R[k]=[\text{Im}(\lambda_k)]^{-1}=\left[-2\lambda\sin {(2\pi\beta k)}\;\sinh{\alpha}\right]^{-1}.
\end{align*}
After including these grounded capacitors and the resistors in Eq.~\eqref{eq:L1}, $\mathcal{L}$ assumes,
\begin{equation}
    \mathcal{L}(\omega_R)=-j\begin{pmatrix}
    -\omega_R C[1]+\frac{j}{R[1]} & \omega_R(C-C') & 0 & \cdots\\
    \omega_R(C+C') & -\omega_R C[2]+\frac{j}{R[2]} &  \omega_R(C-C') & 0 & \cdots \\
    \vdots & \vdots & \vdots & \vdots & \vdots\\
    \cdots 0 & \omega_R(C+C') & -\omega_R C[N+1]+\frac{j}{R[N+1]} & \omega_R(C+C') & 0 \cdots \\
    \vdots & \vdots & \vdots & \vdots & \vdots\\
    \cdots & 0 & \omega_R(C-C') & -\omega_R C[2N]+\frac{j}{R[2N]} & \omega_R(C+C')\\
     & \cdots & 0 & \omega_R(C-C') & -\omega_R C[2N+1]+\frac{j}{R[2N+1]}
    \end{pmatrix}.
    \label{eq:L2}
\end{equation}
\end{widetext}
Thus, Eqs.~\eqref{eq:L1} and \eqref{eq:L2} represent the Laplacians for non-reciprocal TECs in the absence and presence of the complex QP disorder, respectively.

\section{\label{a2}Theoretical time evolution of the TEC}
Earlier, we have highlighted the difficulty of visualizing a specific eigenstate of the Laplacian, $\mathcal{L}$, as a measurable VP.
This challenge arises because every node in the TEC network must be excited via a precisely calibrated current source.
For example, to measure the $k^{\text{th}}$ eigenstate of $\mathcal{L}$, denoted as $V_k$, the amplitude of the current source at each node must satisfy the equation,
\begin{equation*}
I = \mathcal{L}V_k = \zeta_kV_k,
\end{equation*}
where $\zeta_k$ is the $k^{\text{th}}$ eigenvalue of $\mathcal{L}$.
To overcome this challenge, a current source, $I(t)$, can be applied at any node, and the VP $(V(t))$ can be measured at a later time.
Similar to Eq.~\eqref{eq:a_q1}, $V(t)$ can also be expanded as a linear combination of the eigenvectors of $\mathcal{L}$, with time-dependent coefficients $a_k(t)$, 
\begin{equation}
V(t) = \sum_{k=1} a_k(t)V_k. \label{eq:v}
\end{equation}
We assume that $V_k$ forms a complete orthonormal basis for $\mathcal{L}$ with $V_a^{\dagger}V_b = \delta_{ab}$.
Substituting these into Kirchhoff's law leads to
\begin{equation}
I(t) = \mathcal{L}V(t) = \sum_{k=1} a_k(t)\zeta_k V_k;\quad a_k(t) = \frac{V_k^{\dagger}I(t)}{\zeta_k}.
\label{eq:a}
\end{equation}
Thus, the coefficients $a_k(t)$ can be evaluated at any time $t$, provided the input current $I(t)$ is uniquely defined.
The expression for $a_k(t)$ in Eq.~\eqref{eq:a} becomes directly comparable to the coefficients $a_q(t)$ in Eq.~\eqref{eq:a_q2} only when $I(t)$ takes the form of a delta-type excitation, that is, a sharply peaked current pulse with very high amplitude at $t = 0$ and vanishingly short duration.
\begin{figure}[b]
    \centering
    \includegraphics[width=\linewidth]{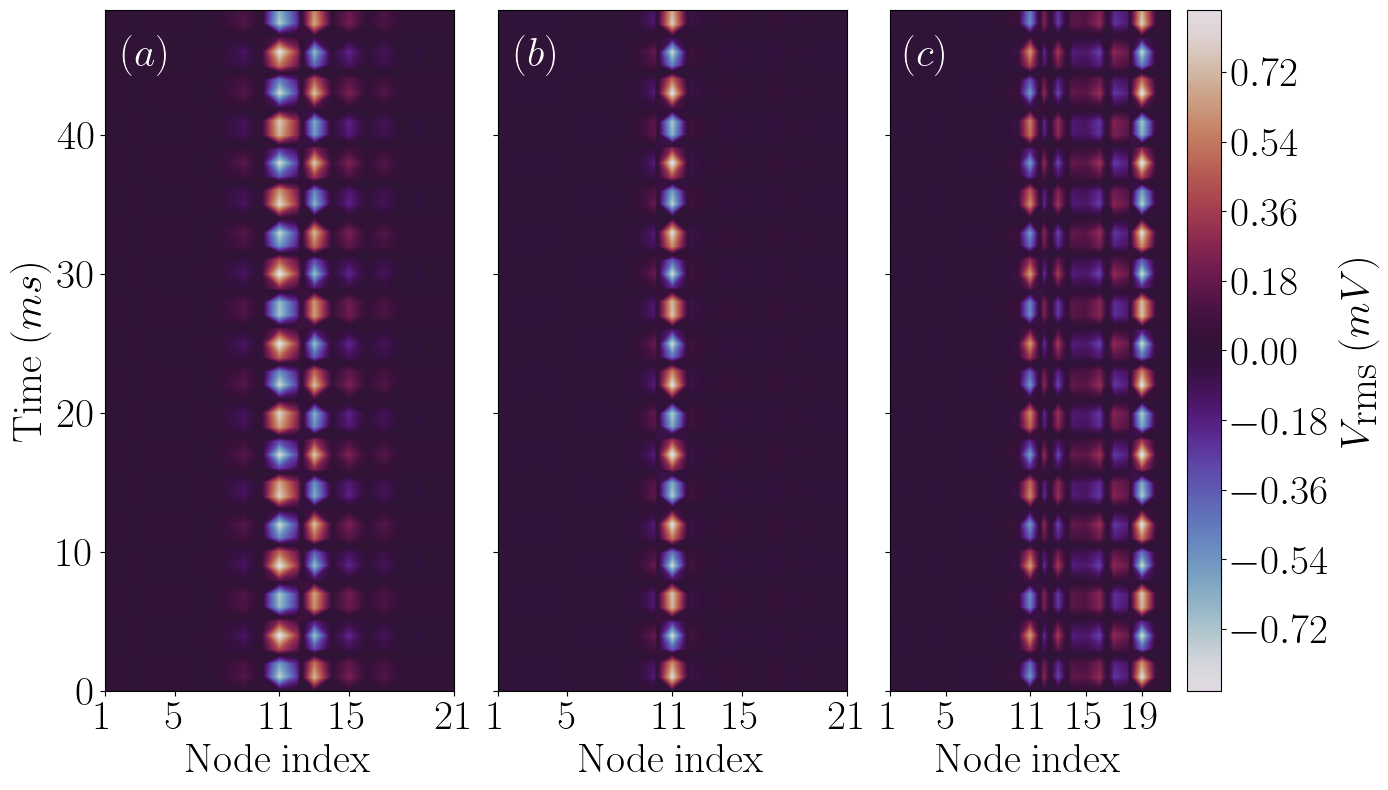}
    \caption{(a) The NHSE is evident as the $V(t)$ localizes at the interface, specifically at $11^{\text{th}}$ node. (b) When $\alpha = \alpha_c$, all the eigenstates of the Laplacian $\mathcal{L}$ transform from the skin states to the AL states, yet the VP remains localized at the interface. (c) Finally, for a larger value of $\alpha$, the localization shifts towards the excitation node, which is the $18^{\text{th}}$ node.}
    \label{fig:7}
\end{figure}
Only under this condition does the time evolution of the VP in Eq.~\eqref{eq:v} accurately correspond to the evolution of the wavefunction $\ket{\Psi(x,t)}$ described in Eq.~\eqref{eq:psi}.
However, to study the dynamical behavior of the TEC, we resort to a specific case and excite the $18^{\text{th}}$ node with a current source, $I(t) = \sin {\omega_R t}$, having an amplitude of $1$ mA and $\omega_R$ representing the resonant frequency of the circuit.
By employing Eqs.~\eqref{eq:v} and \eqref{eq:a}, we obtain $V(t)$, which is then represented as a colormap in Fig.~\ref{fig:7}.
Figs.~\ref{fig:7}(a) and (b) illustrate the scenario for $\alpha \le \alpha_c$, where the NHSE still dominates over AL in the TEC.
This behavior mirrors that of the NH AA model in Figs.~\ref{fig:2}(a) and (b), as the excitation remains localized at the interface (the $11^{\text{th}}$ node).
However, compared to Fig.~\ref{fig:2}(c) for the case of the TB model, no NH jumps are observed in the TEC, as seen in Fig.~\ref{fig:7}(c).
This difference arises because the input current $I(t)$ in the circuit is a sinusoidal function of time rather than a delta-type excitation.
Instead, the VP settles at the $19^{\text{th}}$ node, located in the vicinity of the excitation $(18^{\text{th}})$ node.
\bibliography{ref5}

\begin{thebibliography}{69}
\bibitem [{QP()}]{QP}%
  \BibitemOpen
  \href@noop {} {}\bibinfo {howpublished} {In usual notation, the localization-delocalization transition occurs at $\lambda=2t$. However, we have absorbed a factor of $2$ in the QP potential in Eq. ($1$).}\BibitemShut {Stop}%
\end{thebibliography}

\end{document}